\documentclass[useAMS,usenatbib]{mn2e}
\usepackage{amssymb}
\usepackage{bm}
\usepackage{amsfonts}
\usepackage{amsmath}
\usepackage{graphicx}
\usepackage{subfigure}

\title[Quasar Polarization and Balmer Edge Feature]{Quasars' Optical Polarization and Balmer Edge Feature Revealed by Ultra-Violet, and Polarized Visible to Near Infrared Emissions}
\author[R. Hu \& S. N. Zhang]{Renyu Hu $^{1}$, Shuang-Nan Zhang $^{2,\ 3}$\thanks{E-mail:
zhangsn@ihep.ac.cn}\\
$^{1}$ Department of Earth, Atmospheric and Planetary Sciences,
Massachusetts Institute of Technology, Cambridge, MA 02139, USA\\
$^{2}$ Key Laboratory of Particle Astrophysics, Institute of High Energy Physics, Chinese Academy of Sciences, Beijing 100049, China\\
$^{3}$ Space Science Division, National Astronomical Observatories of China, Chinese Academy of Sciences,
Beijing 100012, China}

\begin{document}

\date{Accepted date. Received  date}

\pagerange{\pageref{firstpage}--\pageref{lastpage}} \pubyear{2011}

\maketitle

\label{firstpage}

\begin{abstract}
Polarized emission from a quasar is produced by wavelength-independent electron scattering surrounding its
accretion disc, and thus avoid the contamination from its host galaxy and reveal the intrinsic emission spectrum
of the accretion disc. Ultra-violet (UV) emission from a quasar is normally free from the contamination from its
host galaxy. Polarization fraction of the quasar's disc emission can therefore be determined by comparing total
UV emission with polarized visible to near-infrared (NIR) emission; and the resulting continuum spectrum from UV
to infrared can reveal the theoretically expected Balmer edge absorption feature. We fit the polarized spectra
in visible and NIR bands together with the total UV spectra of two type-1 quasars (3C 95, 4C 09.72), to an
extended geometrically thin and optically thick accretion disc model. In addition to the standard model, we
include the Balmer edge absorption due to co-rotational neutral gas on a narrow annulus of the accretion disc.
We find that the extended thin accretion disc model provides adequate description on the continuum spectra of
the two quasars from UV to NIR wavelengths. A Monte-Carlo-Markov-Chain fitting to the continuum spectra is able
to well constrain the true polarization fraction of the disk emission, which allows the Balmer edge feature to
be completely revealed from polarized visible to UV continua. The Balmer edge feature is prominent in both
quasars' spectra, and is significantly broadened due to the orbital motion of gas in the accretion disc. The
broadening of the Balmer edge feature is therefore related to the quasar's inclination. This work proves the
concept of determining quasar's inclination from the Balmer edge feature in their continuum spectra.
\end{abstract}


\section{Introduction}

Quasars are the most luminous objects in the Universe. The release of gravitational energy as material is
accreted onto the supermassive black hole (SMBH) heats the accretion disc and generates multi-color thermal
emission from ultraviolet (UV) to infrared (IR) wavelengths (e.g., Shakura \& Sunyaev 1973; Novikov \& Thorne
1973; Lynden-Bell \& Pringle 1974; Laor \& Netzer 1989; Laor et al., 1990; Hubeny et al. 2000). However, the
total emission of a quasar at visible and near-IR (NIR) wavelengths is mixed by emissions from both its
accretion disc surrounding its central SMBH and its host galaxy, which complicates observations of intrinsic
emission of the quasars' accretion disc.  For example, the optically thick and geometrically thin ``standard"
disc model predicts the emission spectrum in the optical to NIR should follow $F_{\nu}\propto\nu^{\alpha}$,
where $F_{\nu}$ is the flux per frequency, $\nu$ is the frequency and $\alpha=1/3$ (e.g., Shakura \& Sunyaev
1973; Lynden-Bell \& Pringle 1974). However many quasar spectra in the optical to NIR have $\alpha<-0.2$
(Neugebauer et al. 1987; Cristiani \& Vio 1990; Francis et al. 1991; Zheng et al. 1997), consistent with
significant host galaxy contaminations which are increasingly important at longer wavelengths.

Polarized spectra of quasars at visible and NIR wavelengths may reveal the emissions of their accretion discs,
because polarization is expected to come from the electron scattering inside the Broad Line Region (BLR) and the
electron scattering is wavelength independent. Kishimoto et al. (2008) reported polarized spectra of six quasars
in the rest wavelength ranging from 0.2 - 2 $\mu$m. These quasars have $\sim$ 1$\%$ of polarization in optical
bands, less than 1$\%$ of polarization in NIR continuum, and little polarization in their emission lines. Indeed
Kishimoto et al. (2008) found that the polarized NIR spectra manifest $\alpha=1/3$, expected from the standard
disc model. We emphasize two important facts regarding quasars' polarization at visible and NIR wavelengths:
first, the alignment of optical polarization with the radio structure of many quasars suggests that the emission
is polarized not in the disc atmosphere but in an equatorial scattering region surrounding the disc (e.g.,
Stockman \& Angel 1979; Smith et al. 2005). Indeed, intrinsic polarization produced in the disc atmosphere is
likely to be damped by Faraday rotation (e.g., Agol \& Blaes 1996). Second, the observed polarization percentage
in the optical to NIR bands may not be the polarization percentage of the intrinsic disc emission, or the
polarization percentage of the equatorial scattering region, due to possible contaminations from the quasar's
host galaxy.

The standard thin disc model predicts a ``bump" shape from UV to visible wavelengths, which is produced by the
emission from the inner disc region and is determined by the combination of the mass of the central supermassive
black hole (BH) and the inner disc boundary radius. The UV continuum is relatively free from the host galaxy's
contamination, and thus should reflect the emergent spectra of the accretion disc and allow to probe the inner
part of the accretion disc. Two quasars of the Kishimoto (2008) sample (3C 95, 4C 09.72) have been observed with
the {\it Hubble Space Telescope} (HST) and have published UV spectra (Bahcall et al. 1993; Marziani et al. 1996;
Evans \& Koratkar 2004). Combining the UV continuum and the polarized spectra in visible and NIR wavelengths,
therefore, will provide powerful spectral diagnosis of the nature of the two quasars' supermassive BHs and
accretion discs. It has also been reported that the quasar continuum spectra in wavelengths shorter than 120 nm
can be described by a simple power law with index $\alpha_{\rm EUV}=-1.76\pm0.12$, which suggests that the
extreme UV (EUV) continuum is probably due to the intergalactic medium photoionized by the integrated radiation
from quasars (Telfer et al. 2002). Therefore the EUV continuum may not reflect the intrinsic emission of a
quasar's accretion disc.

Quasars' continuum spectra from UV to NIR wavelengths may be used to constrain the properties of their central
supermassive BHs, in particular their BH spin, in a similar way to the method used for stellar mass BHs in X-ray
binaries whose BH mass and disc inclination angle ($i$) are known very well for some of them (Zhang et al. 1997;
Shafee et al. 2006). Czerny et al. (2011) used the standard thin accretion disc model to fit several broad-band
photometric points of quasar SDSS J094533.99+100950.1. Assuming a BH mass, this procedure put constraints on its
spin. However, the broad-band photometric fluxes may be contaminated by broad spectral lines, which may affect
the estimates of the BH spin. Davis \& Laor (2011) also used the thin accretion disc model to fit the shape of
optical continuum of a set of 80 PG quasars, and obtained their radiative efficiency and derived their BH spin.
However, both methods had to assume the BH mass and the inclination angle of the accretion disc {\it a priori}.
Various methods can be used to estimate the BH mass in a quasar, but normally within a factor of 2-3, as will be
discussed later in section 4.1.

The inclination of the accretion disc for an individual quasar is normally not well constrained. According to
the unified model of active galactic nuclei, the emission from a quasar's accretion disc is not blocked by its
dusty torus, which implies a disc inclination of less than $\sim60^{\circ}$ (e.g., Antonucci 1993). Advanced
torus model and spectroscopy of quasars in the mid-infrared wavelengths may provide more insights on the
interaction of the disc emission and the torus and better constrain the disc inclination (e.g., Alonso-Herrero
et al. 2011). Another method to determine the disc inclination is to describe not only the spectra but also the
morphology of active galaxies self-consistently (Kacprzak et al. 2011). For radio loud objects, inference of
radio core luminosity may constrain the view angle of the radio source, i.e., the inclination of the accretion
disc (e.g., Wills \& Brotherton 1995). However, the disc inclination has not been estimated from the disc
emission directly. In principle, the luminosity of the accretion disc scales as $\cos(i)$, so the inclination
may be constrained by photometry if the BH mass and accretion rate can be independently estimated (e.g. Czerny
et al. 2011). This method so-far has large uncertainty mainly due to the large uncertainty of BH mass.

The comparison between a quasar's UV emission and its polarized visible emission can be complicated by the
possible Balmer edge absorption. It has been reported that the polarized flux of the two quasars manifest a
discontinuity in the slope at the wavelengths shorter than $\sim$400 nm (Kishimoto et al. 2004). The feature is
mostly interpreted as the buried Balmer edge of the intrinsic spectra of the quasar UV/optical continuum, or the
``Big Blue Bump" (BBB) emission. The edge absorption feature is due to the bound-free opacities in the disc
atmosphere, and indeed indicates the thermal and optically thick nature of the continuum. However, the origin of
the Balmer edge absorption is still unknown. An order-of-magnitude estimate of the broadening of the Balmer edge
indicates that the broadening is consistent with the orbital motion of the corresponding disc annuli responsible
for emission around 400 nm (Kishimoto et al. 2004).

In this work, we use the extended standard thin accretion disc model to fit the polarized spectra in the visible
and near-infrared bands and the total spectra in the UV band of 3C~95 and 4C~09.72, after removing prominent
emission and absorption features; the extension includes the Balmer edge absorption of an optically thin layer
of neutral gas co-rotating with the accretion disc. We use the Monte-Carlo-Markov-Chain (MCMC) method to fully
explore the multi-dimensional parameter space of BH mass, spin, accretion rate, disc inclination and
polarization. The wide spectral coverage of the total UV emission and the polarized spectra in visible and NIR
bands provides a complete description of the accretion disc emission, which allows us to determine of physical
properties of the BH and the accretion disc. In particular, the true polarization fraction of the disc emission
can be well constrained from the spectral fitting, which leads to a continuum spectrum from UV to NIR. The
spectral fitting also constrains the broadening of the Balmer edge, which is constroled by the disc inclination.

\section{Standard Accretion Disc Model with Balmer Edge}

We intend to describe the polarized spectra in visible and NIR wavelengths as well as the total spectra in UV
wavelengths by the standard thin disc model (Shakura \& Sunyaev 1973; Lynden-Bell \& Pringle 1974). As an
extension to the standard model, we include the Balmer edge absorption of the neutral gas that is assumed to be
co-rotating with the accretion disc. We do not intend to model in detail the atmosphere of quasars' accretion
discs (e.g., Loar \& Netzer 1989; Loar et al. 1990; Hubeny et al. 2000) due to the extensive computation burden;
our main goal is to determine the BH parameters from quasars' continuum.

According to the standard thin disc model, its temperature of the accretion disc follows (Krolik 1998; Hubeny et
al. 2000)
\begin{eqnarray}
&T(x)=\bigg[\frac{3GM_{\rm BH}\dot{M}}{8\pi\sigma r^3}R_R(r)\bigg]^{1/4}\nonumber\\
&=5.9\times10^7\ \eta^{-1/4}\ \big(\frac{L}{L_{\rm EDD}}\big)^{1/4}\nonumber\\
&\times\big(\frac{M_{\rm BH}}{M_{\odot}}\big)^{-1/4}\ x^{-3/4}\ R_{\rm R}(x)^{1/4}\quad ({\rm K}), \label{Temp}
\end{eqnarray}
in which $T$ is the temperature, $G$ is the gravitational constant, $M_{\rm BH}$ is the BH mass, $\dot{M}$ is
the accretion rate, $\sigma$ is the Stefan-Boltzmann constant, $r$ is the radius, $x$ is the radius in the unit
of gravitational radius $r_{\rm g}=\frac{GM_{\rm BH}}{c^2}$, $\eta$ is the radiative efficiency as function of
BH spin $a/M$, $L/L_{\rm EDD}$ is the accretion rate in the unit of Eddington accretion rate, and $R_{\rm R}$ is
the correction term of GR effects analytically expressed as a function of $a/M$ (see Krolik 1998). Analytical
expressions of $\eta$ and $R_{\rm R}$ are given in Appendix (\ref{A1}).

We assume that the Balmer edge absorption comes from a relatively narrow disc annuli for effective temperature
from 8,000 to 20,000 K, based on the consideration of the $n=2$ population of hydrogen. The ionization of
hydrogen depends on the number density in the photosphere, which in turn depends on the torque of the disc, an
observational unknown. With a sensible range of the torque value, a realistic disc vertical structure model
(Laor \& Netzer 1989), and the local thermal equilibrium of a pure hydrogen plasma, we find that the $n=2$
population concentrates in a relatively narrow range whose temperature is  from 8,000 to 20,000 K. A large
electron density from ionization of metals, and a range of other parameters might affect the temperature range
of $n=2$ population; we have performed a sensitivity study by letting the Balmer edge absorption to occur in
annuli of $T=8,000\sim15,000-25,000$ K. The frequency-dependent form of the Balmer edge feature is,
\begin{equation}
\begin{cases}
\tau(\nu)=0\ , \quad (\nu<\nu_c)\\
\tau(\nu)=\frac{\tau_0}{\mu}(\nu/\nu_c)^{-2.67}\ , \quad (\nu>\nu_c)
\end{cases}
\end{equation}
where $\tau_0$ is the characteristic optical depth due to the common photoelectric absorption, $\nu_c$ is the
threshold frequency corresponding to $\sim364.6$ nm, and $\mu\equiv\cos(i)$ in which $i$ is the inclination
angle of the accretion disc. We assume $\tau_0$ to be a constant in the Balmer absorption annuli and zero
elsewhere for simplicity. Given $L$, $M_{\rm BH}$, and $a/M$, the range of radii of Balmer absorption annuli
$x_0$ can be found by letting $T=8,000\sim20,000$ K in Eq.~(\ref{Temp}).

The inclination of the accretion disc has three effects: local limb darkening, projection of the disc on the sky
plane, and the relativistic doppler effect due to the orbital motion. Omitting the general relativistic effect
of photon propagation, the asymmetric broadening factor due to the orbital motion is derived as
\begin{eqnarray}
S(\nu,\nu_0) = \frac{1}{\pi}\frac{\nu^3}{\nu_0^4\gamma}\frac{1}{\sqrt{\beta_1^2-(1-\nu/\nu_0\gamma)^2)}} \ ,\nonumber\\
\nu_0\gamma(1-\beta_1) < \nu < \nu_0\gamma(1+\beta_1) \ , \label{broad}
\end{eqnarray}
where $\nu$ is the observed frequency, $\nu_0$ is the emission frequency in the rest frame of the rotating gas,
$\gamma_0\equiv 1/\sqrt{1-\beta_{0}^{2}}$, $\beta_1$ is the relativistic factor corrected for the inclination as
$\beta_1=\beta_0\sin(i)$, $\beta_0\equiv v_0/c$, and $v_0$ is the local Keplerian velocity of the disc where the
Balmer absorption takes place. Here we see that the broadening of the Balmer edge feature sensitively depends on
the disc's inclination. In Newtonian regime, $\beta=x^{-1/2}$. In the Kerr spacetime, $\beta$ is an analytical
function of $x$ and $a/M$ (see Krolik 1998 for detail formulation). We verify that when $x$ is large, the
functional form of $\beta$ in the Kerr spacetime approaches to that of the Newtonian regime.

The emerging spectrum from an optically thick disc with an optically thin layer of absorbing neutral gas is then
described by
\begin{equation}
\nu L_{\nu}=\nu\int_{x_{\rm in}}^{x_{\rm out}} f(x, \nu) 2\pi x dx \times\frac{3}{4}(\mu+\frac{2}{3})\mu \
,\label{Disk}
\end{equation}
where the local limb darkening and projection of the disc plane are considered as in Krolik (1998). The inner
edge of the accretion disc ($x_{\rm in}$) is assumed to be located at the inner most circular orbit of the
spinning BH, as a function of $a/M$. The analytical expression of $x_{\rm in}$ is given in Appendix (\ref{A1}).
In our calculation we choose $x_{\rm out}=1000$, large enough to compute the NIR spectra up to 2 $\mu$m. $f(x,
\nu)$, the emission flux from an annuli between $x$ and $x+dx$, can be computed from the following integration
\begin{equation}
f(x, \nu) = \int_{\nu/\gamma(1+\beta_1)}^{\nu/\gamma(1-\beta_1)}
B(T(x),\nu_0)\exp(-\tau(\nu_0))S(\nu,\nu_0)d\nu_0, \label{Conv}
\end{equation}
where $B(T,\nu)$ is the Planck function. The observed polarized spectrum in the optical and NIR wavelengths
($\nu F_{\nu, \rm OIR}$) and the UV continuum ($\nu F_{\nu, \rm UV}$) are then
\begin{equation}
\nu F_{\nu, \rm OIR}=P \frac{\nu L_{\nu}}{D^2}\ ,\quad \nu F_{\nu, \rm UV}=\frac{\nu L_{\nu}}{D^2}\ ,
\end{equation}
where $P$ is the polarization fraction, and $D$ is the luminosity distance. The luminosity distance to a source
is calculated from the redshift $z$ with the standard flat cosmology model of $(\Omega_{\rm M}=0.3,\ \Omega_{\rm
\Lambda}=0.7)$ and Hubble constant $H_0=72$ km s$^{-1}$ Mpc$^{-1}$.

In summary, our model computes the continuum spectrum from UV to NIR of the thin accretion disc with the effect
of co-rotating optically thin Blamer edge absorption. The model involves 6 independent parameters: BH mass
($M_{\rm BH}$), BH spin ($a/M$), accretion rate ($L$), polarization ($P$), inclination ($i$), and optical depth
of the Balmer absorption ($\tau_0$). In this work we use the kinetics of BLR and the velocity dispersion to
determine the BH mass, which decreases the number of free parameters to 5. We further define the ``effective"
accretion rate as $L'\equiv L \times\frac{3}{4}(\mu+\frac{2}{3})\mu$ according to Equation (\ref{Disk}), in
order to remove the correlation between the accretion rate and the inclination. Finally, $a/M$, $L'$, $P$, $i$
and $\tau_0$ are our five independent fitting parameters. We emphasize that the formulation here is a simplified
treatment of quasars' accretion disc emission continuum. This formulation is computationally efficient, which
allows extensive exploration of the parameter space with the Monte-Carlo-Markov-Chain method.

\section{Data Selection}

We obtain the total-light and polarized spectra of Q0144-3938, 3C~95, CTS A09.36, 4C~09.72, PKS 2310-322 and Ton
202 in the visible and NIR wavelengths from Kishimoto et al. (2008). The optical spectropolarimetry was taken
with the instrument FORS1 mounted on the Very Large Telescope (VLT) UT2 and the Low Resolution Imaging
Spectrograph (LRIS) on the Keck-I telescope (Kishimoto et al. 2004; Kishimoto et al. 2008). 3C~95 and 4C~09.72
were observed with the VLT in September 2002. Q0144-3938, CTS A09.36, PKS 2310-322 were observed with the VLT in
September 2005. Ton 202 was observed with the Low Resolution Imaging Spectrograph (LRIS) on the Keck-I telescope
on May 4, 2003.

The NIR broad-band imaging polarimetry of the six quasars was obtained with the instruments UFTI (with the
polarimetry module IRPOL2) and ISAAC which are mounted on the United Kingdom Infrared Telescope (UKIRT) and VLT
UT1, respectively. Ton 202 was observed with the UKIRT on January 16, 2001 (Kishimoto et al. 2005). 3C~95 and
4C~09.72 were observed with the UKIRT in the fall of 2006. Q0144-3938, CTS A09.36, PKS 2310-322 were observed
with the VLT on August 21, 2007.

Among the sample of the six quasars, 3C~95 and 4C~09.72 have reduced spectra in the near UV wavelengths
available for scientific interpretations. The total-light spectra of 3C~95 and 4C~09.72 in the UV wavelength
were taken by the HST Faint Object Spectrograph (FOS) and available from the archive of the HST High-Level
Science Products (Evans \& Koratkar 2004). The spectrum of 3C~95 was taken on December 21, 1991 with a total
exposure time of 4454 s. The spectrum of 4C~09.72 was taken on October 12, 1992 with a total exposure time of
4751 s and on July 7, 1993 with an exposure time of 480 s, respectively. The UV and optical spectra are
recalibrated uniformly with all other HST pres-COSTAR (Corrective Optics Space Telescope Axial Replacement) FOS
spectrophotometric data for AGNs and quasars, using the up-to-date algorithms and calibration data (Evans \&
Koratkar 2004). Where possible, multiple observations are combined to produce a single spectrum for each object
with the highest possible signal-to-noise ratio and covering the widest wavelength range. We retrieve the
recalibrated spectra of 3C~95 and 4C~09.72 in 1140-3300 $\AA$ in this work.

The retrieved UV spectra have many prominent emission and absorption lines, and have the spectral resolution of
$\sim$1 $\AA$. The UV spectra have been processed in the following steps before taken into the spectral
fittings. First, we remove the absorption lines in the UV spectra. For 3C~95, we use the complete sample of
absorption lines provided by Bahcall et al. (1993), which contains the central wavelength in the observation
frame and FWHM of each absorption line. We assume a Gaussian profile for each absorption line with three
parameters: central wavelength, FWHM, and strength. We take the central wavelength and the FWHM from Bahcall et
al. (1993), and fit the UV spectrum to determine the strength. We list in Table \ref{Labs} central wavelengths,
FWHM, and strengths of all absorption lines. For 4C~09.72, because no absorption line list is available from the
literature, we do not perform the removal of absorption lines; because the absorption lines are all very narrow
with FWHM less than 3 $\AA$, they are unlikely to affect the fitting to the continuum.
\begin{table}
 \caption{Absorption lines removed from the UV spectra of 3C~95. The wavelength is in the observation frame.}
 \begin{tabular}{llccl}
  \hline
  $\lambda$ & FWHM  & $\lambda F_{\lambda} \times 10^{11}$               & Line  \\
  (nm)      & (nm)  & erg s$^{-1}$ cm$^{-2}$) & \\
  \hline \hline
1649.02   & 1.51  & 0.30   & Lyman $\alpha$ \\ \hline 1742.76   & 1.86  & 0.18   & Lyman $\alpha$ \\ \hline
1749.42   & 1.65  & 0.32   & Lyman $\alpha$ \\ \hline 1890.00   & 2.08  & 0.35   & Si IV \\ \hline 1909.82   &
1.88  & 0.56   & Lyman $\alpha$ \\ \hline 1959.37   & 1.59  & 0.58   & Lyman $\alpha$ \\ \hline 2025.72   & 1.94
& 0.25   & Zn II (Mg II) \\ \hline 2099.70   & 1.84  & 0.53   & C IV \\ \hline 2103.38   & 1.80  & 0.43   & C IV
\\ \hline 2314.49   & 4.64  & 0.14   &  \\ \hline 2382.53   & 2.04  & 0.29   & Fe II \\ \hline 2586.82   & 2.04
& 0.18   & Fe II \\ \hline 2600.20   & 2.13  & 0.31   & Fe II \\ \hline 2755.34   & 2.04  & 0.17   &  \\ \hline
2796.25   & 2.23  & 0.35   & Mg II \\ \hline 2803.27   & 2.55  & 0.25   & Mg II \\ \hline
2852.95   & 2.04  & 0.19   & Mg I \\
  \hline \hline
 \end{tabular}
\label{Labs}
\end{table}

\begin{figure}
\includegraphics[width=84mm]{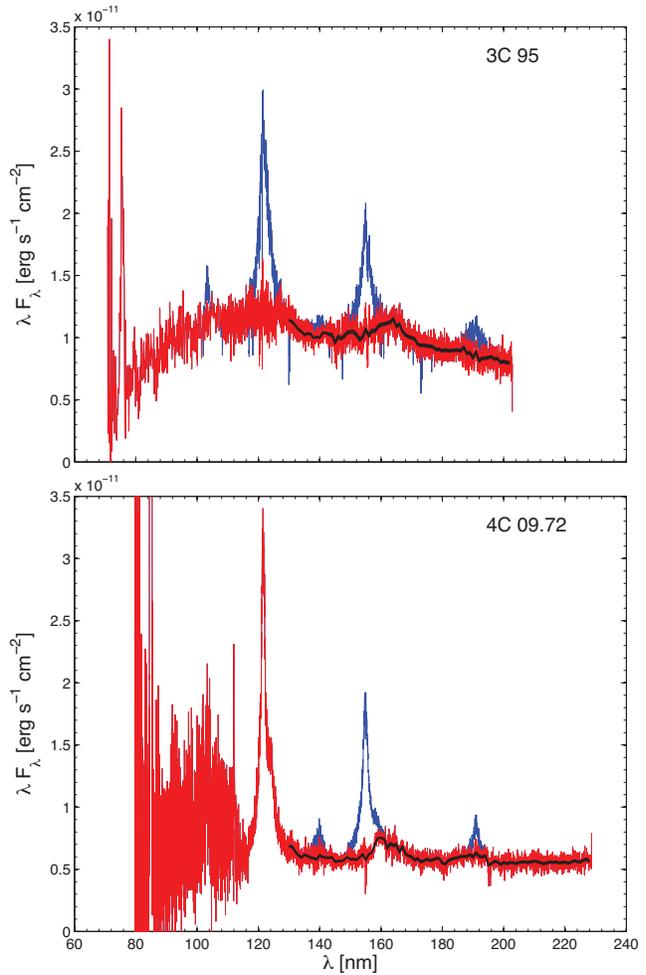}
\caption{Pre-fitting processes of the UV spectra of 3C~95 and 4C~09.72. Blue curves show the original spectra
retrieved from the archive of the HST High-Level Science Products with wavelengths in the rest frame. Red curves
show the spectra after the emission and absorption lines removed. Black curves show the spectra with a linear
wavelength spacing of 1 nm.} \label{UVFig}
\end{figure}

Second, we remove the emission lines in the UV spectra. The extraction of emission lines, especially the
prominent broad lines, are complicated. Marziani et al. (1996) discussed the profiles of emission lines of 3C~95
and 4C~09.72 in details. In this work, however, we only focus on the continuum. Hence, we try to remove the
emission lines in a simple way. We use the line identifications and central wavelengths given by the HST
High-Level Science Products and remove each emission line by several trial Gaussian profiles. Note that there is
no spectral fitting performed here. We test each of the emission lines with several combinations of FWHM and
line strength, assuming the Gaussian profiles, and determine the best line removal by manual inspection. The
Gaussian profiles retracted from the UV spectra are tabulated in Table \ref{Lemi}. We do not attempt to remove
the iron emission lines. For both 3C~95 and 4C~09.72, the Fe II and Fe III pseudocontinua are much weaker than
the C IV emission (Marziani et al. 1996). Also, we only use the UV continua for the rest-frame wavelengths
shorter than $\sim200$ nm; in this wavelength range the effect of iron lines to our spectral fitting is minor
(e.g., Vestergaard \& Wilkes 2001). Finally, for 3C~95 and 4C~09.72, we dismiss the spectrum with wavelength
shorter than 1300 $\AA$ in the rest frame because of the prominent Galactic Lyman-$\alpha$ emission and large
statistical errors (e.g. Evans \& Koratkar 2004), the non-thermal component due to Compton scattering in the
accretion disc atmosphere (e.g., Hubeny et al. 2001), and possible intergalactic contamination (Telfer et al.
2002). The UV continuum spectra after the absorption and emission line removal are shown in Figure \ref{UVFig}.

\begin{table}
 \caption{Emission lines removed from the UV spectra of 3C~95 and 4C~09.72. The wavelength is in the observation frame.}
 \begin{tabular}{llccl}
  \hline
  Object & $\lambda$ & FWHM & $\lambda F_{\lambda} \times 10^{11}$ & Line  \\
  & (nm) & (nm) & (erg s$^{-1}$ cm$^{-2}$) & \\
  \hline
\hline
  3C~95 & 1670.9   & 25    & 0.30   & Ly$\beta$, O VI \\
\hline
  3C~95 & 1964.9   & 70    & 0.80   & Ly$\alpha$ \\
\hline
  3C~95 & 1964.9   & 30    & 0.50   & Ly$\alpha$ \\
\hline
  3C~95 & 2004.5   & 60    & 0.30   & N V \\
\hline
  3C~95 & 2257.4   & 50    & 0.10   & Si IV \\
\hline
  3C~95 & 2504.4   & 90    & 0.50   & C IV \\
\hline
  3C~95 & 2504.4   & 35    & 0.30   & C IV \\
\hline
  3C~95 & 3084.9   & 80    & 0.20   & C III] \\
\hline \hline
4C~09.72 & 2001.9     & 40    & 0.15   & Si IV \\
\hline
4C~09.72 & 2220.9     & 80    & 0.50   & C IV \\
\hline
4C~09.72 & 2220.9     & 20    & 0.75   & C IV \\
\hline
4C~09.72 & 2735.7     & 40    & 0.20   & C III] \\
  \hline
\hline
 \end{tabular}
\label{Lemi}
\end{table}

Finally, we transform the spectra into the rest frame by multiplying the flux $F_{\lambda}$ by $(1+z)$ and
present the flux in terms of $\lambda F_{\lambda}$ (see blue curves in Figure \ref{UVFig}). For the fitting
purpose we also bin the UV spectra with a linear wavelength spacing of 1 nm, as shown in Figure \ref{UVFig}. The
uncertainties of the flux are combined by the standard error propagation method.

\section{Data Analysis}

\subsection{Black Hole Mass Determination}

The most direct measurements of black masses come from the kinematics of the broad line regions (BLRs) of AGNs,
i.e., the emission-line width $\delta V$ and the size of BLR given by reverberation mapping (Blandford \& Mckee
1982; Peterson 1993), assuming that the line emitting gas in the BLR is varialized. The virial mass can then be
expressed as
\begin{equation}
M_{\rm BH-RM}=\frac{fR\delta V^2}{G} ,
\end{equation}
where $f$ is a factor depending on the structure, kinematics and orientation of the BLR (Peterson et al. 2004).
On average, the factor $f$ can be taken as $<f>=5.5$ if the line width is taken as the line dispersion
$\sigma_{\rm line}$. For a distant AGN the reverberation measurement is not possible; then the BLR size comes
from the radius-luminosity (R-L) relationship (the relationship between the radius of BLR and the optical
luminosity) calibrated at low redshift (Kaspi et al. 2005). The removal of the host-galaxy starlight
contribution improves the relationship significantly, and the relationship manifests as $r\propto L^{1/2}$
(Bentz et al. 2006).

The BH mass determined this way has been found to be tightly correlated with the velocity dispersion of the
bulge or spheroid $\sigma_*$ (the so-called $M_{\rm BH}-\sigma_*$ relation; Ferrarese \& Merritt 2000; Tremaine
et al. 2002). In this work, the $M_{\rm BH}-\sigma_*$ relation we used is determined from the
reverberation-mapped virial BH mass, which is provided by Onken et al. (2004). In general, the FWHM of the [O
III] $\lambda$5007 emission line can be used as a proxy for $\sigma_*$ (Nelson \& Whittle 1996; Nelson 2000).

Although we only perform the spectral fitting to the continuums of 3C~95 and 4C~09.72 in the following, we
determine the BH masses for all six quasars in the sample from the BLR kinematics and compare them with the
$M_{\rm BH}-\sigma_*$ relation. In this way we can verify the reliability of the BH mass determination. We
obtain the broad H$\beta$ line width by interpolating a linear continuum between continuum windows on either
side of the line, and calculate the line's mean squared dispersion defined as the second moment of the line
profile (Peterson et al. 2004). In practice, the narrow-line residuals influence significantly the determination
of the broad-line width. Therefore we remove the narrow-line component carefully using the adjacent [O III]
$\lambda$5007 line as a template. The obtained line widths and BH masses are tabulated in column 1 of Table
\ref{MassT}. The BH masses determined by such method have typical errors by a factor of 2, i.e., 0.3 dex
(Peterson et al. 2004).

\begin{figure}
\includegraphics[width=84mm]{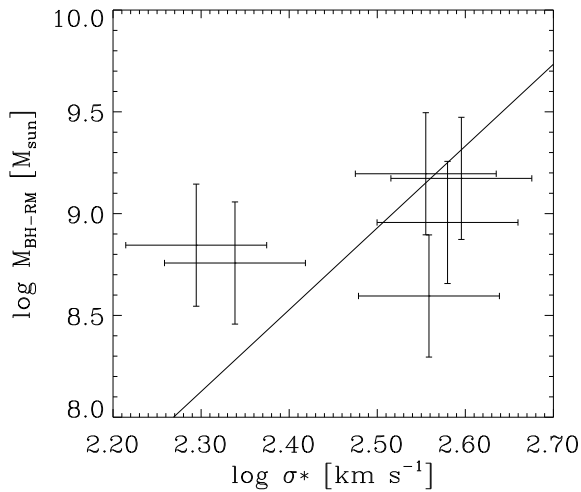}
\caption{The virial black hole masses $M_{\rm BH}$ and the velocity dispersion $\sigma_*$ on $M_{\rm
BH}-\sigma_*$ relations. The solid line shows the $M_{\rm BH}-\sigma_*$ relation by Tremaine et al. (2002). The
uncertainty of $M_{\rm BH}$ is within a factor of 2 (Peterson et al. 2004), and the uncertainty of ${\rm
log}(\sigma_*)$ is 0.07 dex (Tremaine et al. 2002).} \label{MsigmaF}
\end{figure}

Using the line dispersion of [O III] $\lambda$5007 as the probe of the velocity dispersion of an AGN's host
galaxy's bulge, we can plot the obtained BH masses with respect to the established $M_{\rm BH}-\sigma_*$
relation (Tremaine et al. 2002), as in Figure \ref{MsigmaF}. Using the Nukers' estimate, we find the reduced
$\chi^2$ is 1.2, if errors for $M_{\rm BH-RM}$ and $\sigma_*$ are taken as 0.3 dex (Peterson et al. 2004) and
0.07 dex (Tremaine et al. 2002), respectively. $M_{\rm BH-RM}$ measures the virial mass the BH, and $\sigma_*$
indicates the dispersion mass of BH. The consistence of these independent measurements as shown in Figure
\ref{MsigmaF} allows us to conclude that BH mass estimates with both methods are reliable for these sources. To
further improve the accuracy of these BH's mass estimates, we combine the BH mass estimates from both methods;
the errors for the combined BH masses are thus 0.21 dex. In Table 1, we list the observation data used and the
BH masses determined from the reverberation mapping (denoted as $M_{\rm BH-RM}$), the velocity dispersion
(denoted as $M_{{\rm BH}-\sigma}$), and the combined kinematic measurements $\log M_{\rm BH-KM}=0.5\times(\log
M_{\rm BH-RM}+\log M_{{\rm BH}-\sigma})$.

\begin{table*}
 \caption{Quasar black hole mass estimates}
 \label{MassT}
 \begin{tabular}{lccccccccccccc}
  \hline
  Quasar name & H$\beta$ width & BLR size$^1$ & $M_{\rm BH-RM}$$^2$ & $\sigma_*$$^3$ & $M_{{\rm BH}-\sigma}$$^4$ & $M_{\rm BH-KM}$$^5$& z \\
 & (km s$^{-1}$) & (Light days) & ($10^8$M$_{\odot}$) & (km s$^{-1}$) & ($10^8$M$_{\odot}$) & ($10^8$M$_{\odot}$)\\
  \hline
  Q0144-3938 & 2570 & 55.8 & 3.94 & 362 & 14.7 & 7.6 & 0.244 \\
  \hline
  3C~95 & 2846 & 171.5 & 14.9 & 394 & 20.6 & 17.5 & 0.616 \\
  \hline
  CTS A09.36 & 3124 & 54.8 & 5.72 & 218 & 1.9 & 3.3 & 0.310 \\
  \hline
  4C~09.72 & 2855 & 179.8 & 15.7 & 359 & 14.2 & 14.9 & 0.433 \\
  \hline
  PKS~2810-322 & 3048 & 70.4 & 7.00 & 197 & 1.3 & 3.0 & 0.337 \\
  \hline
  Ton~202 & 3306 & 77.4 & 9.05 & 380 & 17.8 & 12.7 & 0.366 \\
  \hline
 \end{tabular}
 \begin{enumerate}
 \item[1] The r-L relationship of Kaspi et al. (2005) is applied.\\
 \item[2] Estimated from H$\beta$ width and BLR size, with a 1-$\sigma$ error of 0.3 dec. The factor $f$ is taken as $f=5.5$.\\
 \item[3] The velocity dispersion is taken as the line dispersion of [O III]
 $\lambda$5007.\\
 \item[4] Estimated from $\sigma_*$, with a 1-$\sigma$ error of 0.3 dec.\\
 \item[5] Geometrical average between $M_{\rm BH-RM}$ and $M_{{\rm BH}-\sigma}$, i.e., $\log M_{\rm BH-KM}=0.5\times(\log M_{\rm BH-RM}+\log M_{{\rm BH}-\sigma})$, with a 1-$\sigma$ error of 0.21 dec.
 \end{enumerate}
\end{table*}

\subsection{Spectral fitting}

We fit the two observed quasar continuum spectra (3C~95, 4C~09.72) to our extended standard think disc model
presented above, with the polarized spectra in the visible and NIR wavelengths, and the total spectra in the UV
wavelengths. The polarized spectra are bridged with the total light spectra by the polarization fraction of the
observed accretion disc emission. The 5 fitting parameters are BH spin $a/M$, accretion rate $L'$, polarization
fraction $P$ inclination $i$, and Balmer edge optical depth $\tau_0$. In the fitting, we use the BH mass
estimate $M_{\rm BH-KM}$ in table \ref{MassT}.

We use the Monte Carlo Markov Chain (MCMC) method for the spectral fitting to explore the whole parameter space.
Due to the multi-parameter nature of the problem, the MCMC method not only provides possibility of fast
computation, but also gives the posterior probability distribution of all parameters. We use the adaptive
Metropolis-Hastings algorithms implemented by Haario et al. (2006).

The convergence of the Markov chain is examined as follows. We start from two randomly selected parameter sets
and generate two Markov chains. The second halves of both chains are mixed together and the so-called
``potential scale reduction" ($R$) is computed for each estimated parameter (Gelman \& Rubin 1992); $R$ declines
to unity as the length of Markov chain goes to infinity. In this work we require $R<1.005$ for all estimated
parameters as the criterium for the convergence of Markov chain. We find that this requirement corresponds to a
Markov chain length of 1 millions, so for each quasar at least 2 millions continuum spectra have been generated
to be compared with the data.

\begin{figure*}
\includegraphics[width=168mm]{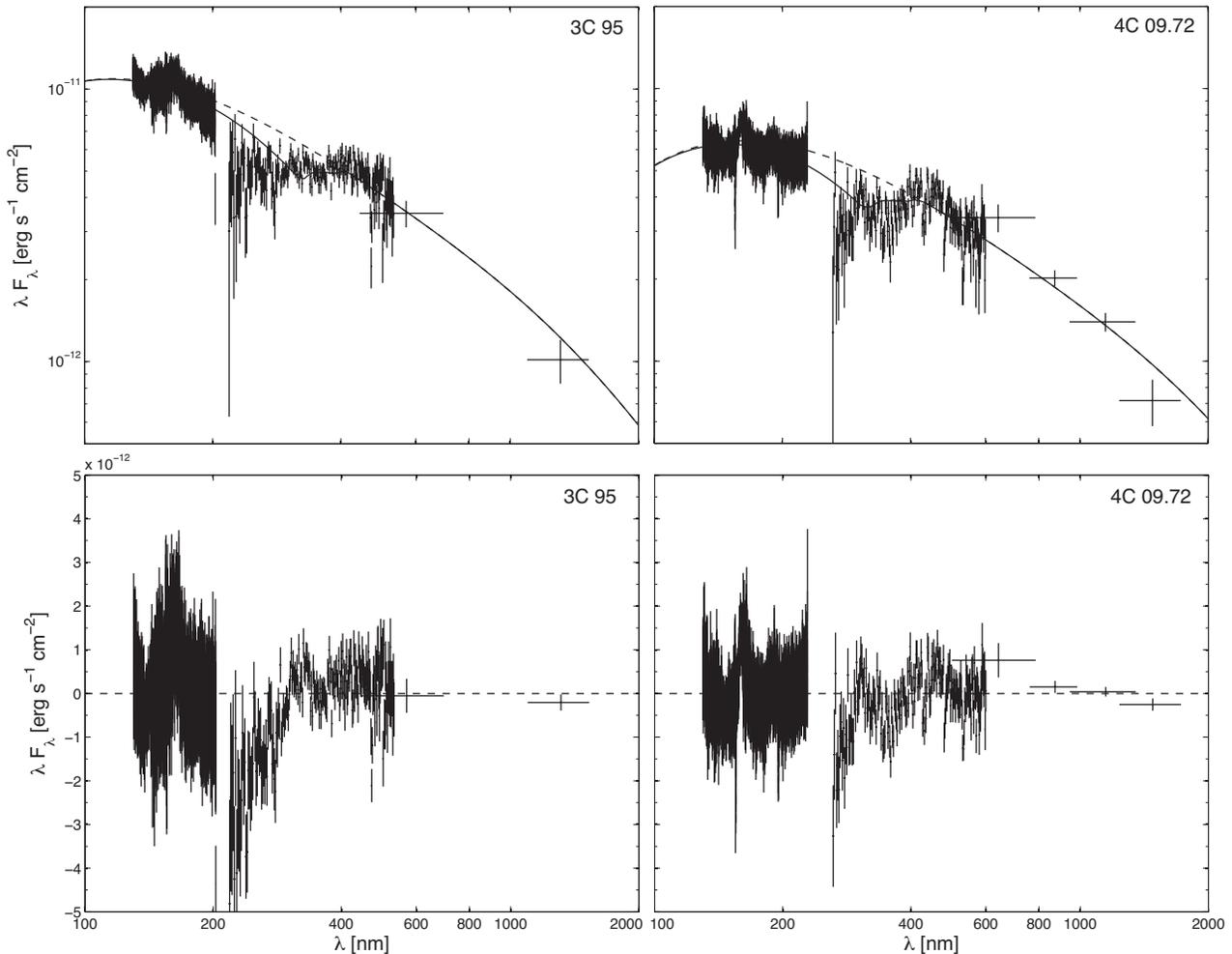}
\caption{Best fitted model spectra of 3C~95 and 4C~09.72. The upper panels show the observed spectra and the
fitting results. The observed polarized spectra have been uniformly multiplied by a factor of $1/P$ to be
compared with the UV total light continuum. The solid lines are the best fitted model spectra with the
parameters tabulated in Table \ref{Fit}. The dashed lines are calculated with the same parameters but without
the Balmer-edge absorption ($\tau_0=0$). The lower panels show the fitting residuals. Note that the vertical
axis of the lower panels is in the linear scale. We see that the Balmer edge absorption is clearly revealed in
the quasars' polarized continuum spectra.} \label{Fit1}
\end{figure*}

\begin{figure*}
\includegraphics[width=168mm]{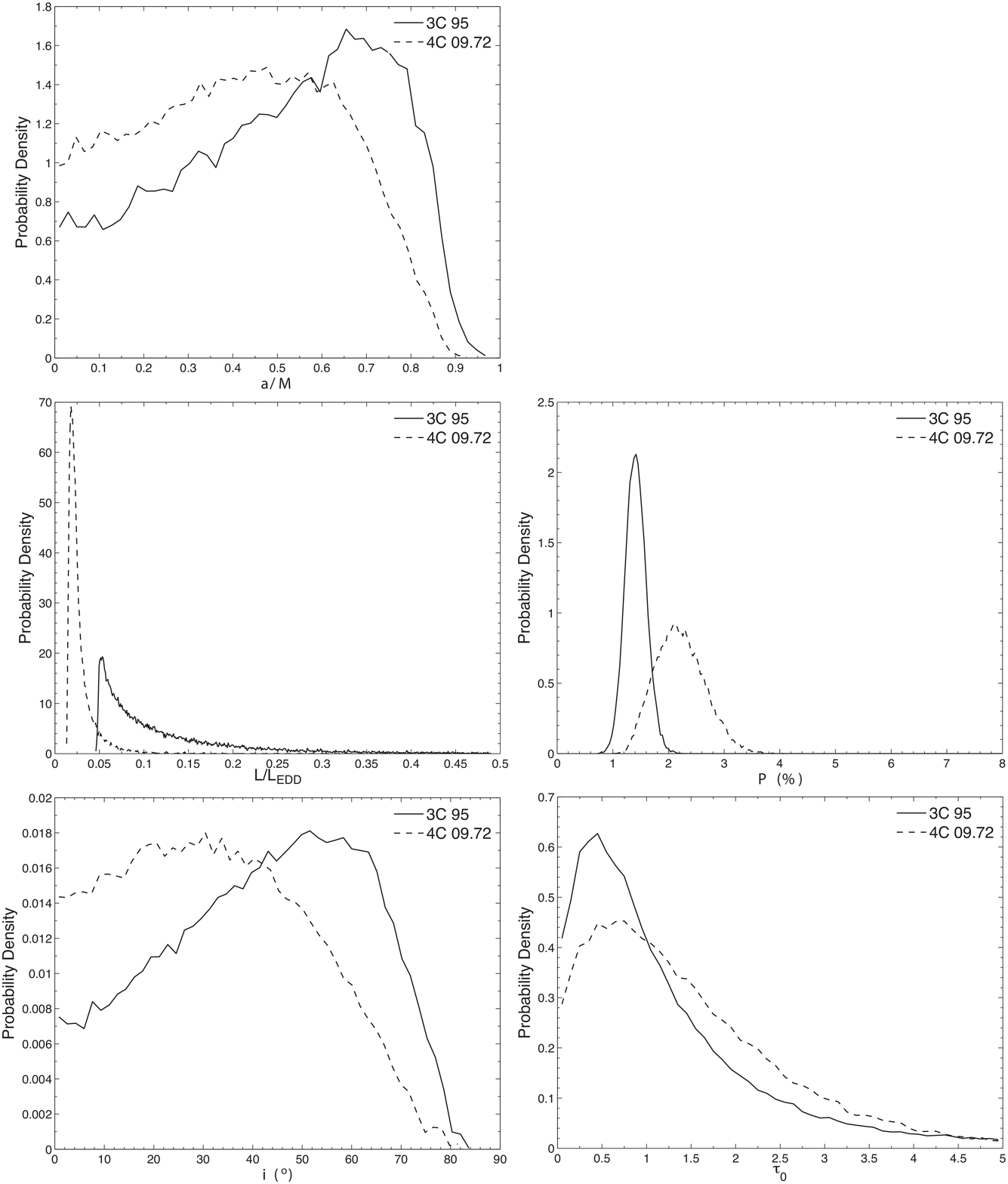}
\caption{Posterior probability densities of physical parameters of 3C~95 and 4C~09.72. For each object, the
probability densities are computed from the MCMC chain of the last 2 millions iterations to minimize the effect
of initial conditions. The accretion rate ($L/L_{\rm EDD}$) is computed from the effective accretion rate ($L'$)
and the inclination ($i$) of the fitting results. We see that UV continuum and polarized optical continuum can
well constrain the accretion rate and the polarization, and the balmer edges revealed by polarized optical
continuum provides information of the inclination. } \label{Fit3}
\end{figure*}

The fitting results are tabulated in Table \ref{Fit} and the best fitted model spectra are shown in Figure
\ref{Fit1} as solid lines. As comparison, we also plot the corresponding model spectra without the Balmer-edge
absorption as dashed lines. The probability density of each parameter from the MCMC simulation is shown in
Figure \ref{Fit3}. The optimal value ($X_{\rm O}$) of each parameter is taken as the best-fitted value of the
parameters (i.e., the parameter set that provides the smallest $\chi^2$ in the Markov chain). Due to the
asymmetry of these probability distributions, the lower ($X_{\rm L}$) and upper ($X_{\rm U}$) boundaries at 95\%
confidence level of each parameter is determined in the following way:
\begin{equation}
\int_{X_{\rm L}}^{X_{\rm O}}p(X)dX/\int_{0}^{X_{\rm O}}p(X)dX=0.95, \end{equation} and
\begin{equation}\int_{X_{\rm O}}^{X_{\rm U}}p(X)dX/\int_{X_{\rm O}}^{\infty}p(X)dX=0.95,\end{equation}
where $p(X)$ is the probability density of the parameter $X$. The correlations among the five fitting parameters
of 3C~95 and 4C~09.72 are shown in Figure \ref{3Cpair} and Figure \ref{4Cpair}, respectively. We have provided
both the best fitted values and the most probably values for the parameters in Table \ref{Fit}. The
discrepancies between the best-fitted values and the most-probable values can be quite large, for those
parameters that are poorly constrained by the data (e.g., $a/M$, $i$). Nevertheless, for all parameters the
best-fitted values and the most-probable values lie within their 1-$\sigma$ ranges.

\begin{table*}
 \caption{Physical parameters of 3C~95 and 4C~09.72 estimated by fitting their continuum spectra.
Five parameters are estimated from the spectral fitting: black hole spin $a/M$, accretion rate $L'/L_{\rm
EDD}\equiv L/L_{\rm EDD} \times\frac{3}{4}(\mu+\frac{2}{3})\mu$, polarization $P$, disc inclination $i$ and
Balmer edge optical depth $\tau_0$. The spectral fitting is performed with MCMC simulations that generate chains
to the convergence. The black hole masses are estimated by kinetic measurements ($M_{\rm BH-KM}$ in Table
\ref{MassT}). The first line of each object lists the best fitted results, defined as the parameter set that
provides the smallest $\chi^2$. The second line of each object lists the most probable parameter values, defined
as the peak value in the posterior probability distribution. The third line of each object lists the lower and
upper 95\% boundary values of parameter estimates on both sides of the best-fitted values for
$T=8,000\sim20,000$~K, using Eqs. (8) and (9), respectively. The fourth line lists the parameter boundary values
estimated from a separate set of MCMC simulations, in which we have assumed the Balmer edge absorption occurs in
the disk's annuli of $T=8,000\sim15,000$~K. Similary, the fifth line lists the parameter boundary values
estimated with $T=8,000\sim25,000$~K. We see that except for $\tau_0$, the spectral fitting does not depend on
the specific choice of the Balmer edge temperature range. }
 \label{Fit}
 \begin{tabular}{l|ccccc|ccc}
  \hline
    & $a/M$ & $L'/L_{\rm EDD}$ & P (\%) & $i$ $(^{\circ})$ & $\tau_0$ & $\chi^2/dof$  &  $L/L_{\rm EDD}$ \\
  \hline
  3C~95 (Best-fitted)       & 0.030 & 0.070 & 1.48 & 70.4 & 0.32 & 4237.8/2637 & 0.28 \\
  Peak  & 0.65  & 0.067 & 1.37 & 51.6 & 0.45 &  & 0.11 \\
  $T=(8-20)\times 10^3$~K & 0.0035 -- 0.84 & 0.061 -- 0.080 & 1.08 -- 1.83 & 6.6 -- 79.6 & 0.035 -- 4.80  &  & 0.049 -- 0.70 \\
  $T=(8-15)\times 10^3$~K & 0.0054 -- 0.82 & 0.062 -- 0.084 & 1.10 -- 1.86 & 8.1 -- 82.7 & 0.055 -- 9.51  &  & 0.050 -- 1.11 \\
  $T=(8-25)\times 10^3$~K & 0.0045 -- 0.84 & 0.061 -- 0.081 & 1.10 -- 1.82 & 6.8 -- 80.0 & 0.021 -- 1.60  &  & 0.050 -- 0.75 \\
  \hline
  4C~09.72 (Best-fitted)  & 0.0049 & 0.021 & 2.46 & 67.1 & 0.34 & 7678.7/3295 & 0.067 \\
  Peak  & 0.48   & 0.020 & 2.09 & 30.4 & 0.75 &  & 0.020 \\
  $T=(8-20)\times 10^3$~K & 0.0024 -- 0.76 & 0.018 -- 0.027 & 1.49 -- 3.25 & 3.5 -- 78.3 & 0.035 -- 4.04 & & 0.014 -- 0.20 \\
  $T=(8-15)\times 10^3$~K & 0.0034 -- 0.74 & 0.018 -- 0.024 & 1.60 -- 3.12 & 3.9 -- 75.3 & 0.135 -- 9.56 & & 0.015 -- 0.14 \\
  $T=(8-25)\times 10^3$~K & 0.0034 -- 0.77 & 0.018 -- 0.027 & 1.47 -- 3.09 & 3.3 -- 74.8 & 0.036 -- 1.96 & & 0.015 -- 0.15 \\
  \hline
 \end{tabular}
\end{table*}

\section{Results}

\subsection{Polarization of Accretion Disc Emission}

The continuum spectra of 3C~95 and 4C~09.72 from UV to NIR wavelengths can be sufficiently described by the
standard accretion disc model with a prominent Balmer-edge absorption, as shown in Figure \ref{Fit1}. The peak
emissions of both 3C~95 and 4C~09.72 are at the wavelengths of 100$\sim$200 nm, which have been observed by the
HST. The observed UV continuum spectra of the two quasars are indeed the optically thick thermal radiation from
the inner part of the accretion disc.  The polarized spectra from visible to NIR wavelengths are
wavelength-independently scaled spectra of the disc's thermal emission; therefore the polarized spectra from
visible to NIR wavelengths can be bridged with the total UV continua by adjusting the polarization fraction. The
resulting continuum spectra from UV to NIR represent the thermal emission from a wide radius range of the
quasars' accretion discs.

\begin{figure*}
\includegraphics[width=168mm]{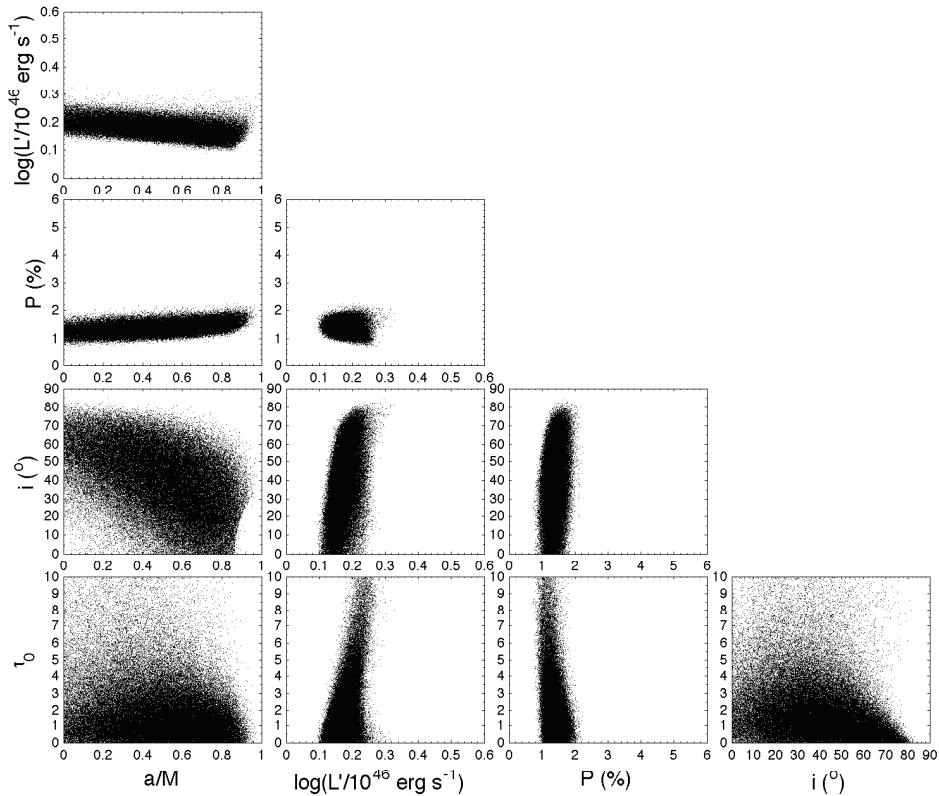}
\caption{Correlation of the five fitting parameters of 3C~95. Each panel shows the joint posterior probability
distribution between a pair of parameters with each dot corresponding to a parameter set in the MCMC chain of
the second half of the 2 million iterations. } \label{3Cpair}
\end{figure*}

\begin{figure*}
\includegraphics[width=168mm]{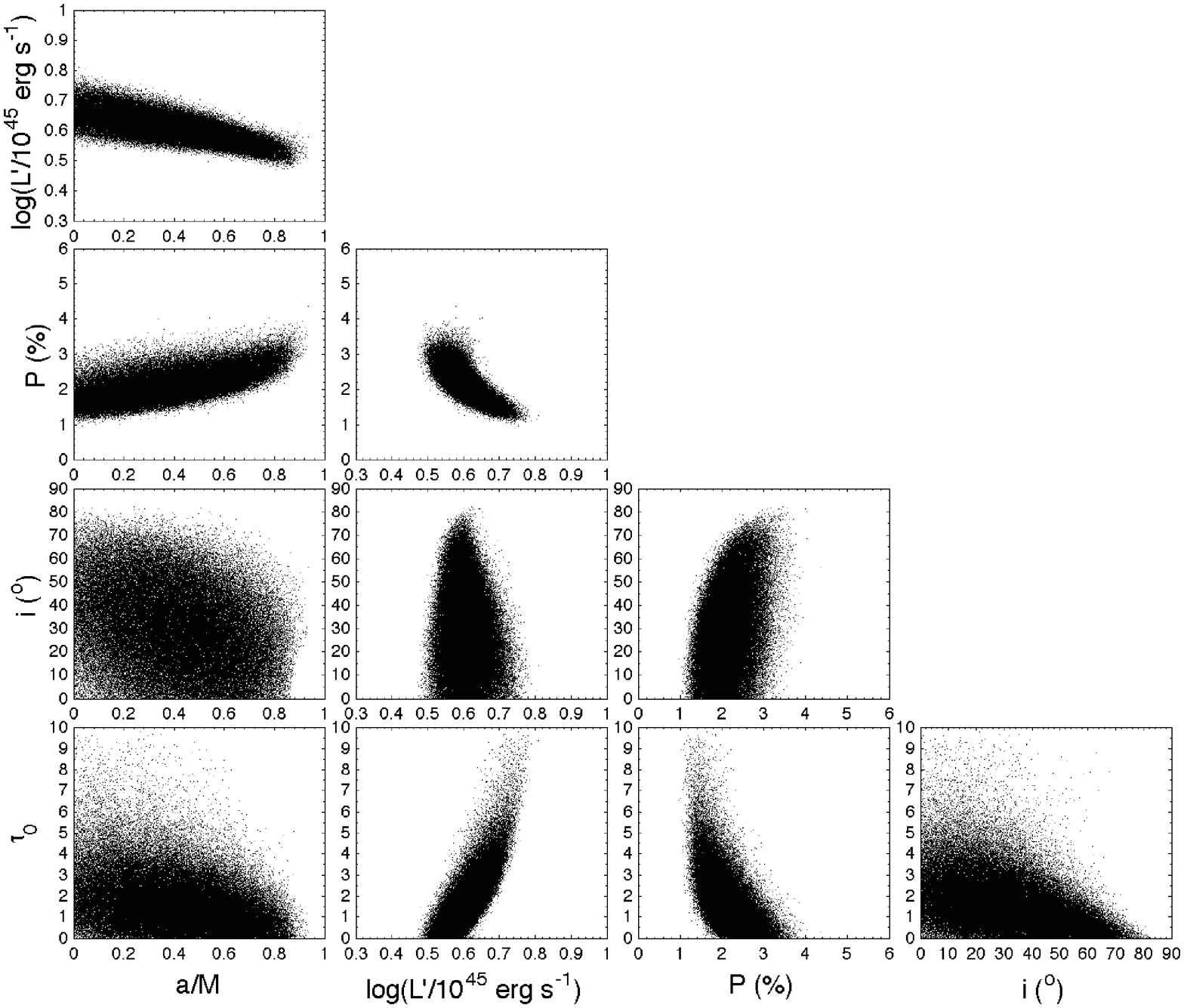}
\caption{Correlation of the five fitting parameters of 4C~09.72. Each panel shows the joint posterior
probability distribution between a pair of parameters with each dot corresponding to a parameter set in the MCMC
chain of the second half of the 2 million iterations.} \label{4Cpair}
\end{figure*}

Optical polarization of accretion disc emission can be determined by bridging the polarized visible-IR emission
to the total emission at UV wavelengths. Both total emission and polarized emission, although at different
wavelengths, follow the same multi-color thermal emission spectrum; hence the polarization fraction can be well
constrained by spectral fitting of the continuum thermal emission of accretion discs (see Figure \ref{Fit3}).
The uncertainty of the optical polarization is gaussian and independent from other BH or accretion disc
properties (see Figure \ref{3Cpair} and Figure \ref{4Cpair}). We emphasize that the polarization derived in this
paper is the true polarization of the accretion disc emission, and we expect the true polarization to be larger
than the observed polarization due to host galaxy contamination. The observed polarization fraction of both
3C~95 and 4C~09.72 is $\sim1.0-1.5\%$ at wavelengths of $400\sim500$ nm and gradually decreases to
$\sim0.5-1.0\%$ at $250$ nm (Schmidt \& Smith 2000; Kishimoto et al. 2008). These observations are consistent
with our spectral fitting results, although we do not use total light in visible wavelengths directly. In
comparison with the true polarization obtained from spectral fitting, we find that for 3C~95, there is little
contamination from its host galaxy at $400\sim500$ nm, but the contamination increases significantly for shorter
wavelengths; for 4C~09.72, host galaxy contamination is significant for all visible and NIR wavelengths.

Spectral fitting to the accretion disc continuum from UV to NIR may constrain the BH and accretion disc
properties. The disc's accretion rate is tightly constrained by the total luminosity of the accretion disc, a
physical quantity that is derived from the polarized luminosity and the true polarization fraction (see Figure
\ref{Fit3}). The accretion rate is only weakly correlated with the BH spin, as predicted by the standard thin
disc model (see Figure \ref{3Cpair} and Figure \ref{4Cpair}). Also, the estimated accretion rates are
significantly sub-Eddington, well within the range that the standard thin disc model can be applied reliably. We
have also performed continuum spectral fitting for 3C~95 and 4C~09.72 with the BH mass assumed as the fitting
parameter (not shown in the paper). We obtained quantitatively very similar results and the BH mass estimated
from the spectral fitting is consistent with the kinetics estimates in Table \ref{MassT}. This suggests that
their disc continuum spectra do not contain additional information on their BH masses, i.e., their BH mass
uncertainties are dominated by the uncertainties in their kinematic mass estimates.

\subsection{Revealed Balmer Edge Absorption}

Prominent Balmer edge absorption features are found in their continuum spectra, as shown in Figure \ref{Fit1}.
The Balmer edge absorption is due to the bound-free opacities in the accretion disc atmosphere, which can only
be seen in the disc's emission continuum with host galaxy contamination removed. We here confirm that the Balmer
edge absorption for quasars is observable via polarized spectra in the visible wavelengths (i.e., Kishimoto et
al. 2004; 2005; 2008).
 The predicted shape of the Balmer edge absorption combines
the photoelectric absorption profile and the broadening due to the orbital motion. As shown in Figure
\ref{Fit1}, the Balmer edge absorption fits the break of polarized spectra at $\sim400$ nm and connects to the
UV continuum smoothly.

The observed Balmer edge absorption is broadened due to the orbital motion of the accretion disc. Since the
Balmer edge opacity is proportional to the concentration of $n=2$ hydrogen, the Balmer edge absorption feature
in the disc continuum is generated in a narrow annuli of the accretion disc in which the temperature is
$8,000\sim20,000$ K. The inclination of the accretion disc affects significantly the broadening of the Balmer
edge feature, as the broadening mostly depends on the line-of-sight projection of the orbital motion velocity of
the specific Balmer-edge-generating annulus. From the spectral fitting, we find that for one thing, orbital
motion is sufficient to account for the broadening of the revealed Balmer edges for 3C~95 and 4C~09.72 (Figure
\ref{Fit1}); for another, the widths of Balmer edges may constrain the discs' inclination (Figure \ref{Fit3}).
Noticeably, the edge-on geometry is confidently ruled out by fitting emission spectra for 3C~95 and 4C~09.72
(Figure \ref{Fit3}), consistent with them being type-1 quasars. However, the spectral fitting cannot rule out
the face-on geometry (Figure \ref{Fit3}), because of the intrinsic width of the Balmer edge from the
$\nu/\nu_c^{-2.67}$ dependency of the photoelectric absorption cross sections. One might assert that the
most-likely inclination of 3C~95 is larger than that of 4C~09.72; any further constraint of disc inclination by
Balmer edge broadening is impeded by a relatively low signal-to-noise ratio of the Balmer edge feature, and the
uncertainties of Balmer edge optical depths.

Spectral fitting to the accretion disc continuum from UV to NIR can determine the optical depth of the Balmer
edge absorption. As shown in Table \ref{Fit}, the Balmer edge optical depths of 3C~95 and 4C~09.72 are large
enough to create significant spectral features. The upper limit of the optical depth is constrained by the
spectral fitting, indicated by the posterior probability distribution shown in Figure \ref{Fit3}. We caution
that for the Balmer edge optical depth significantly larger than unity the optically thin assumption may break
down and the full radiative-transfer model of the accretion disc atmosphere is needed. After all, the
best-fitted values and the most-probable values of the optical depth for both quasars are found to be consistent
with an optically thin absorption. With the bound-free cross section of hydrogen at the Balmer edge
($\sigma_{\rm bf}=1.3\times10^{-17}$ cm$^{2}$), we estimate the column density of excited ($n=2$) hydrogen atoms
to be $3\times10^{15}$ -- $4\times10^{17}$ cm$^{-2}$ for 3C~95 and $3\times10^{15}$ -- $3\times10^{17}$
cm$^{-2}$ for 4C~09.72, respectively. For local thermal equilibrium the implied neutral hydrogen column density
on the accretion disc by the Balmer edge optical depth is estimated to be $10^{19}\sim10^{21}$ cm$^{-2}$. The
neutral hydrogen column density is comparable to the neutral hydrogen column density of the Milky Way, which is
consistent with the interpretation that the Balmer edge features originate on the quasars' accretion discs.

\section{Conclusion and Discussion}

A quasar's polarized continuum spectrum at visible to NIR wavelengths, together with the total continuum
spectrum at UV wavelengths, provides an unique opportunity to determine the properties of accretion discs around
supermassive BHs. In this work we fitted the spectra of two type-1 quasars (3C~95 and 4C~09.72) to the standard
thin accretion disc model with the Balmer-edge absorption, with the BH masses determined from stellar dynamics
and reverberation measurements. We used the MCMC method to sample a parameter space of BH spin, accretion rate,
polarization, disc inclination and Balmer edge optical depth.

We conclude that the quasar polarized spectra at visible and NIR wavelengths and total spectra in UV wavelengths
can reveal the accretion disc emission, which is indeed the thermal emission of the optically thick and
geometrically thin accretion disc. At the UV wavelengths, the contamination from the host galaxy is negligible
so the total spectra reflects the emission spectra of the accretion disc. At the visible and NIR wavelengths,
the contamination buries the intrinsic accretion disc emission as well as the Balmer-edge feature. However the
contamination does not exist in the polarized spectra. Therefore, the polarized spectra at visible and NIR
wavelengths and total spectra at UV wavelengths allow us to study the entire accretion disc. The true optical
polarization of the accretion disc emission can be well constrained by bridging the total emission at UV and the
polarized emission at visible. We find that 3C~95 and 4C~09.72 have significantly sub-Eddington accretion rates,
and their optical polarization is $1.1-1.8\%$ and $1.5-3.3\%$ (at 95\% confidence level), respectively. The true
optical polarization of accretion disc derived from our continuum fitting is larger than the observed optical
polarization, which indicates host galaxies' contamination in the quasars' total optical continua.

The absorption feature found at the Balmer edge is prominent and significantly broadened. We found that the
Balmer absorption feature for both quasars investigated can be reproduced by assuming optically thin absorbing
$n=2$ hydrogen population co-rotating with the accretion disc at radii with effective temperature between 8,000
and 20,000 K. The Balmer edge feature is found to extend to UV wavelengths, whose width is the result of both
photoelectric absorption cross section and the line-of-sight projection of the orbital motion. The choice of the
temperature range does not significantly affect the spectral fitting. For example, for the upper boundary of
temperature varying from 15,000 to 25,000 K, we did not find any significant corresponding dependency of the
black hole parameters and the true optical polarization (see Table \ref{Fit}). However, we found the best-fitted
Balmer edge optical depth ($\tau_0$) to be strongly dependent on the width of Balmer absorption annulus. The
wider the annulus is, the deeper the Balmer edge appears, and the required $\tau_0$ to fit the data is smaller
(see Table \ref{Fit}). Therefore the estimation of neutral hydrogen on the surface of the accretion disc is also
sensitive to the temperature range. The sensitivity study has suggested that a reasonable uncertainty in the
exact location where the Balmer edge absorption occurs has a major effect in the depth of the Balmer edge
feature, but only has a minor effect on the broadening of the Balmer edge feature. Indeed, the broadening of the
Balmer edge feature depends mostly on the disc's inclination. The broadening of the Balmer edge features
tentatively constrain the disc's inclination of the two quasars.

Our fitting results show that the quasars' BH spin parameters cannot be constrained well from the spectral
fitting up to $\sim100$ nm. As shown in Figure \ref{Fit3} and Table \ref{MassT}, the posterior probability
distribution of the BH spin is very wide. The best-fitted parameters for both 3C~95 and 4C~09.72 indicate
non-rotating BHs; however, the full exploration to the parameter space indicates that a significant range of
$a/M$ between 0 to 1 is acceptable. This is due to the combination of the uncertainties in inclination,
accretion rate and BH mass.

Figure \ref{3Cpair} and \ref{4Cpair} show a weak correlation between $\tau_0$ and $i$. As expected, a larger
$\tau_0$ would produce a deeper absorption edge, which allows the widths of the Balmer edge feature to be better
determined, and the disc's inclination to be better constrained. Indeed, the uncertainties in the observed
Balmer edge depth play a major role in the uncertainty of inclination. For example, the two polarized spectra at
wavelengths shorter than about 300 nm are obviously below the model predictions. The spectral features were
noticed by Kishimoto et al. (2004), who did not make unambiguous identification and interpretation of these
features, but suggested that they might be related to Fe II absorption and possibly the Bowen
resonance-florescence lines. Nevertheless, the relatively lower signal to noise ratios of the measurements of
the polarized spectra at wavelengths shorter than about 300 nm do not seem to affect the fitting results to the
Balmer edge significantly. However, future better observations may allow identification and physical modeling of
these features, which should allow better determination of the Balmer edge.

The foreseeable improvement in tightening the inclination uncertainty is to have better polarized spectral
observations around the Balmer edge in the future. In this study, we find that the Balmer edge optical depth and
the disc inclination are particularly sensitive to the continuum emission at 300 - 400 nm wavelengths. The
current data have low signal-to-noise ratio in this wavelength range; with high signal-to-noise ratio data in
the future, it is plausible to study the visible continuum of the quasar accretion discs, and constrain the
Balmer edge absorption and the disc inclination.  Furthermore, since the uncertainty in inclination angle will
propagate into the uncertainty in accretion rate for a given observed flux, a more accurate inclination will
also result in a more accurate accretion rate. Therefore a combination of a more accurate inclination and a well
observed peak emission will then allow the BH mass and accretion rate determined accurately, leading eventually
to accurate BH spin measurement.

\appendix

\section{Analytical Expressions of $\lowercase{x_{\rm in}}$, $\eta$ and $R_{\rm R}$}

\label{A1}

We provide analytical expressions of the the (dimensionless) inner most circular orbit ($\lowercase{x_{\rm
in}}$), the radiative efficiency ($\eta$) and the GR correction term ($R_{\rm R}$), as functions of the BH spin
($a/M$) (Krolik 1998). The inner most circular orbit ($x_{\rm in}$) is
\begin{equation}
x_{\rm in}=3+z_2-\sqrt{(3-z_1)(3+z_1+2z_2)} \ ,
\end{equation}
for which $z_1$ and $z_2$ can be computed from the BH spin ($a/M$) as
\begin{equation}
z_1=1+[1-(a/M)^2]^{1/3} [(1+a/M)^{1/3}+(1-a/M)^{1/3}] \ ,
\end{equation}
\begin{equation}
z_2=\sqrt{3(a/M)^2+z_1^2} \ .
\end{equation}
The radiative efficiency is related to the inner most circular orbit as
\begin{equation}
\eta = 1-\frac{x_{\rm in}^2-2x_{\rm in}+(a/M) x_{\rm in}^{1/2}}{x_{\rm in}\sqrt{x_{\rm in}^2-3x_{\rm in}+2(a/M)
x_{\rm in}^{1/2}}} \ .
\end{equation}

The GR correction term ($R_{\rm R}$) is a function of radius $x$ as
\begin{equation}
R_{\rm R}(x) = \frac{C(x)}{A(x)} \ ,
\end{equation}
where $A(x)$ is
\begin{equation}
A(x) = 1-\frac{2}{x}+\frac{(a/M)^2}{x^2} \ ,
\end{equation}
and $C(x)$ is
\begin{eqnarray}
C(x)&=&1-\frac{y_{\rm in}}{y}-\frac{3(a/M)}{2y} \log\bigg(\frac{y}{y_{\rm in}}\bigg) \nonumber\\
&&-\frac{3(y_1-a/M)^2}{y y_1 (y_1-y_2) (y_1-y_3)} \log\bigg(\frac{y-y_1}{y_{\rm in}-y_1} \bigg) \nonumber\\
&&-\frac{3(y_2-a/M)^2}{y y_2 (y_2-y_1) (y_2-y_3)} \log\bigg(\frac{y-y_2}{y_{\rm in}-y_2} \bigg) \nonumber\\
&&-\frac{3(y_3-a/M)^2}{y y_3 (y_3-y_2) (y_3-y_1)} \log\bigg(\frac{y-y_3}{y_{\rm in}-y_3} \bigg) \ ,
\end{eqnarray}
in which $y=\sqrt{x}$, $y_{\rm in} = \sqrt{x_{\rm in}}$, and $y_1$, $y_2$, and $y_3$ are three roots of the
following equation:
\begin{equation}
x^3-3x+2a/M=0 \ .
\end{equation}

\section*{Acknowledgments}

We thank Makoto Kishimoto for providing us the electronic version of their published polarized spectra used in
this work. R. Y. H. thanks L. Zhu for helpful discussion. The first referee Robert Antonucci and the second,
anonymous referee are thanked for their constructive criticisms and many helpful comments, which allowed us to
correct and improve the paper substantially. R. Y. H. acknowledges partial funding support by the MIT
Presidential Fellowship. S.N.Z. also acknowledges partial funding support by the National Natural Science
Foundation of China under grant nos. 11133002, 10821061, 10733010, 10725313, and by 973 Program of China under
grant 2009CB824800.

\bsp

\label{lastpage}


\begin{thebibliography}{99}

\bibitem[1996]{Agol1996} Agol, E., Blaes, O., 1996, MNRAS, 282, 965

\bibitem[2011]{Alonso-Herrero2011} Alonso-Herrero A., et al., 2011, ApJ, in press

\bibitem[1993]{Antonucci1993} Antonucci R., 1993, ARA\&A, 31, 473

\bibitem[1993]{Bahcall1993} Bahcall J. N., et al., 1993, ApJS, 87, 1


\bibitem[2006]{Bentz2006} Bentz M. C., Peterson B. M., Pogge R. W.,
Vestergaard M., \& Onken C. A., 2006, ApJ, 644, 133



\bibitem[1982]{Blandford1982} Blandford R. D., \& Mckee C. F., 1982,
ApJ, 255, 419



\bibitem[]{} Chandrasekhar S., 1960, Radiative Transfer, Dover Publications, Inc., New York

\bibitem[\protect\citeauthoryear{Cristiani
\& Vio}{1990}]{1990A&A...227..385C} Cristiani S., Vio R., 1990, A\&A, 227, 385


\bibitem[2011]{Czerny2011} Czerny B., Hryniewicz K., Nikolajuk M., Sadowski A., 2011, astro-ph(1104.2743)

\bibitem[2011]{Davis2011} Davis S. W., Laor A., 2011, ApJ, 728, 98



\bibitem[2004]{Evans2004} Evans I. N., Koratkar A. P., 2004, ApJS, 150, 73

\bibitem[2000]{Ferrarese2000} Ferrarese L., \& Merritt D., 2000, ApJ, 539, L9

\bibitem[\protect\citeauthoryear{Francis et
al.}{1991}]{1991ApJ...373..465F} Francis P.~J., Hewett P.~C., Foltz C.~B., Chaffee F.~H., Weymann R.~J., Morris
S.~L., 1991, ApJ, 373, 465

\bibitem[1992]{Gelman1992} Gelman A., Rubin D.~B., 1992, Statistical Science, 7, 457


\bibitem[2006]{ Haario2006} Haario H., Laine M., Mira A., Saksman E., 2006, Statistics and Computing, 16, 339

\bibitem[2000]{Hubeny2000} Hubeny I., Agol E., Blaes O., \& Krolik J. H., 2000,
ApJ, 533, 710

\bibitem[2001]{Hubeny2001} Hubeny I., Blaes O., Krolik J. H., \& Agol E., 2001, ApJ, 559, 680

\bibitem[2011]{Kacprzak2011} Kacprzak G. G., Churchill C. W., Evans J. L., Murphy M. T., Steidel C. C., 2011, MNRAS, in press

\bibitem[2005]{Kaspi2005} Kaspi S., Maoz D., Netzer H., Peterson B.
M., Vestergaard M., \& Jannuzi B. T., 2005, ApJ, 629, 61

\bibitem[2004]{Kishimoto2004} Kishimoto M., Antonucci R., Boisson C., \& Blaes
O., 2004, MNRAS, 354, 1065

\bibitem[2005]{Kishimoto2005} Kishimoto M., Antonucci R. \& Blaes O., 2005, MNRAS, 364, 640¨C648 (2005).

\bibitem[2008]{Kishimoto2008} Kishimoto M., Antonucci R., Blaes O., Lawrence A., Boisson C., Albrecht M., \& Leipski C., 2008, Nature, 454,
492

\bibitem[1998]{Krolik1998} Krolik J. H., 1998, Active Galactic Nuclei:
From the Central Black Hole to the Galactic Environment, Princeton University Press

\bibitem[1989]{Loar1989} Loar, A., Netzer, H., 1989, MNRAS, 238, 897

\bibitem[1990]{Loar1990} Loar, A., Netzer, H., Piran, T., 1990, MNRAS, 242, 560







\bibitem[1996]{Marziani1996} Marziani P., Sulentic J. W., Dultzin-Hcyan D., Calvani M., \& Moles M., 1996, ApJS, 104, 37.



\bibitem[2000]{Nelson2000} Nelson C. H., 2000, ApJ, 544, L91

\bibitem[1996]{Nelson1996} Nelson C. H., \& Whittle M., 1996, ApJ, 465, 96

\bibitem[\protect\citeauthoryear{Neugebauer et
al.}{1987}]{1987ApJS...63..615N} Neugebauer G., Green R.~F., Matthews K., Schmidt M., Soifer B.~T., Bennett J.,
1987, ApJS, 63, 615

\bibitem[1973]{Novikov1973} Novikov, I., Thorne, K. S., 1973, In Black Holes, p. 422, eds. de Witt, C. \& de Witt, B., Gordon \& Breach, New York

\bibitem[2004]{Onken2004} Onken C. A., Ferrarese L., Merritt D.,
Peterson B. M., Pogge R. W., Vestergaard M., \& Wandel A., 2004, ApJ, 615, 645



\bibitem[1983]{Peterson1993} Peterson B. M., 1993, PSAP, 105, 247

\bibitem[2004]{Peterson2004} Peterson B. M., et al., 2004, ApJ, 613,
682

\bibitem[\protect\citeauthoryear{Pringle
\& Rees}{1972}]{1972A&A....21....1P} Pringle J.~E., Rees M.~J., 1972, A\&A, 21, 1


\bibitem[2000]{Schmidt2000} Schmidt, G. D., Smith, P. S., 2000, ApJ, 545, 117

\bibitem[2009]{Schnittman2009} Schnittman, J. D., Krolik, J. H., 2009, ApJ, 701, 1175

\bibitem[2006]{Shafee2006}Shafee, R., McClintock, J.E., Narayan, R., Davis, S.W., Li, L-X., Remillard, R.A., 2006, ApJ, 636, L113

\bibitem[\protect\citeauthoryear{Shakura
\& Sunyaev}{1973}]{1973A&A....24..337S} Shakura N.~I., Sunyaev R.~A., 1973, A\&A, 24, 337

\bibitem[2005]{Smith2005} Smith, J. E., Robinson, A., Young, S., Axon, D. J., Corbett, E. A., 2005, MNRAS, 359, 846

\bibitem[1979]{Stockman1979}, Stockman, H. S., Angel, J. R. P., Miley, G. K., 1979, ApJ, 227, L55

\bibitem[2002]{Telfer2002} Telfer R. C., Zheng W., Kriss G. A., Davidsen A. F., 2002, ApJ, 565, 773

\bibitem[2002]{Tremaine2002} Tremaine S. et al., 2002, ApJ, 574, 740

%



\bibitem[]{} Vestergaard, M., Wilkes, B.J., 2001, ApJS, 134, 1

\bibitem[]{} Wills, B. J., Brotherton, M. S., ApJ, 1995, 448, L81

\bibitem[]{} Zhang S. N., Cui W., Chen W., ApJ, 1997, 482, L155-L158

\bibitem[\protect\citeauthoryear{Zheng et al.}{1997}]{1997ApJ...475..469Z}
Zheng W., Kriss G.~A., Telfer R.~C., Grimes J.~P., Davidsen A.~F., 1997, ApJ, 475, 469


\end{thebibliography}
\end{document}